\documentclass[prl,twocolumn,groupedaddress,showpacs]{revtex4}
\usepackage{graphicx}
\begin{document}


\title{Full Frequency Back-Action Spectrum of a Single Electron
Transistor during Qubit read-out}


\author{G\"oran Johansson}
\email[]{tfygj@fy.calmers.se}
\author{Andreas K\"ack}
\author{G\"oran Wendin}
\affiliation{Microtechnology Center at Chalmers MC2, Department of Microelectronics and Nanoscience, Chalmers University of Tecnology and G\"oteborg University, S-412 96, G\"oteborg, Sweden}


\date{\today}

\begin{abstract}
We calculate the spectral density of voltage fluctuations in a
Single Electron Transistor (SET), biased to operate in a transport mode
where tunneling events are correlated due to Coulomb interaction.
The whole spectrum from low frequency shot noise to quantum noise at
frequencies comparable to the SET charging energy $(E_{C}/\hbar)$
is considered.
We discuss the back-action during read-out
of a charge qubit and conclude that single-shot read-out
is possible using the Radio-Frequency SET.
\end{abstract}
\pacs{03.67.Lx 42.50.Lc 73.23.Hk 85.25.Na}

\maketitle

The Single Electron Transistor (SET) has recently been suggested
as a read-out device for solid-state charge qubits 
\cite{Aassime,Kane,e_on_He,Averin,MakhlinPRL}.
Presenting new state of the art figures for sensitivity in
charge measurements, Aassime et al.~\cite{Aassime} made probable
that the Radio-Frequency SET~\cite{RFSET} (RF-SET)
can be used for single shot read-out
of the Single Cooper-pair Box (SCB) qubit~\cite{SCB_Bouchiat,Nakamura}.
This is possible if the measurement time $t_{ms}$ needed to resolve
the two states of the qubit is much shorter than
the time $t_{mix}$ required to destroy the initial qubit state
because of back-action due to voltage fluctuations on the SET.

The mixing rate $1/t_{mix}$ is proportional to the noise
at the frequency $\omega$ corresponding to the energy splitting between
the qubit states $\Delta E$.
Moreover, in order to have well defined
charge states to achieve single shot read-out,
$\Delta E$ should be comparable to the qubit charging energy $E_{qb}$.
The ratio $E_{qb}/E_{C}$ may be varied between approximately
1/10 and 10 in order to optimize the performance of the system.
 
The voltage fluctuations on the SET island have previously
been studied in the experimentally accessible low frequency limit
($\hbar\omega \ll E_{C}$). The noise is here fully understood
both in the classical transport regime of sequential tunneling,
described by a master equation with frequency independent
tunneling rates~\cite{Korotkov_first,Hanke,Korotkov_spectral},
as well as in the Coulomb blockade cotunneling regime~\cite{Averin}.
In Ref.~\cite{Aassime} the back-action of the SET was estimated
using simple interpolation between the low frequency shot
noise limit and the high frequency limit of Nyquist noise from two
independent tunnel junctions.

In this paper we evaluate the finite frequency noise
on a fully quantum mechanical basis, taking into account
the frequency dependence of the tunneling rates in the presence
of fluctuations, separating processes that absorb or emit
energy. We present a simple expression for
the voltage noise of a SET biased in the transport mode,
valid in the whole frequency range from low frequency classical
shot noise to high frequency quantum noise. By comparing $t_{mix}$
with $t_{ms}$ for reading out a SCB qubit with the RF-SET
of Ref.~\cite{Aassime}, we conclude that $t_{ms} \ll t_{mix}$, i.e.
single-shot read-out should indeed be possible.

\begin{figure}
\includegraphics[width=5cm]{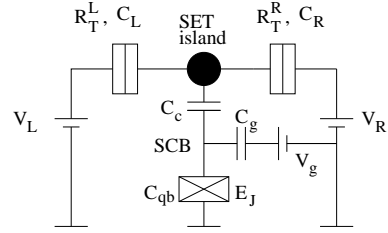}%
\caption{Schematic figure of the SET capacitively coupled to the SCB.}
\label{qnoise_SET_fig}
\end{figure}

Consider a small metallic SET island coupled via low transparency
tunnel barriers to two external leads, and coupled capacitively to 
the SCB, which is controlled by
a gate voltage (see Fig.\ref{qnoise_SET_fig}). First we calculate the voltage
noise on the SET island taking the voltage $V_g$ on the SCB to be
constant. Then the back-action of these fluctuations on the state
of the SCB is estimated. This approach is appropriate in the considered limit
of weak SET-qubit coupling 
($\kappa C_c/C_{qb}, \kappa \ll 1; C_L, C_R \sim C_{qb}$).

We follow the outline of Ref.~\cite{SchoellerPRB} and model
the SET by the Hamiltonian
\begin{equation}
H=H_L+H_R+H_I+V+H_T=H_0+H_T,
\end{equation}
where
\begin{equation}
H_r=\sum_{kn}\epsilon^r_{kn}a^\dagger_{krn}a_{krn},
\qquad
H_I=\sum_{ln}\epsilon_{ln} c^\dagger_{ln} c_{ln}
\end{equation}
describe noninteracting electrons in the 
left/right lead ($H_r, r\in\{L,R\}$) and on the island ($H_I$).
The quantum numbers $n$ denote transverse channels including spin,
and $k, l$ denote momenta. The Coulomb interaction on the
island is described by
\begin{equation}
V(\hat{N})=E_C(\hat{N}-n_x)^2,
\end{equation}
where $\hat{N}$ denotes the excess number operator,
$E_C=e^2/2C$ the charging energy ($C=C_L+C_R+C_c$),
$n_x$ the fractional number of electrons induced by the
external voltages
($n_x$ is the fractional part of $(C_L V_L + C_R V_R + C_c V_g)/e$)
and $e$ the electron charge. The tunneling term is
\begin{equation}
H_T=\sum_{r=L,R}\sum_{kln}(T^{rn}_{kl}\,a^\dagger_{krn}c_{ln}
e^{-i\hat{\Phi}}\,+\,T^{rn*}_{kl}\,c^\dagger_{ln}a_{krn}
e^{i\hat{\Phi}}),
\end{equation}
where the operator $e^{\pm i\hat{\Phi}}$ changes the excess
particle number on the island by $\pm 1$ and $T^{rn}_{kl}$
are the tunneling matrix elements. $\hat{\Phi}$ is the canonical
conjugate to $\hat{N}, ([\hat{\Phi},\hat{N}]=i)$. In this case of
a metallic island, containing a large number of electrons, the charge
degree of freedom $N=0,\pm1,...$ is to a very good approximation
independent of the electron degrees of freedom $l, n$.

The Cooper-pair box
is a small superconducting island coupled to a superconducting
reservoir, via a tunnel junction with
Josephson energy $E_J$ (see Fig.~\ref{qnoise_SET_fig}).
The box is also characterized by its charging energy
$E_{qb}=e^2/2C_{qb}$,
where $C_{qb}$ is the total capacitance of the box.
For $E_J\ll E_{qb}$ two nearby charge states, corresponding to zero
and one excess Cooper pair on the island, can be used as
the $|0\rangle$ and $|1\rangle$ states of the qubit.
The energy splitting $\Delta E$ between these states is
controlled by the gate voltage $V_g$.
%

%
%
For a given charge measurement sensitivity $\delta q$ and
measurement time $t_{ms}$, the uncertainty in charge is given by
$\Delta q=\delta q/\sqrt{t_{ms}}$. In order to separate the two qubit
states we need the two intervals $0\pm\Delta q$
and $2e\pm\Delta q$ not to overlap, which defines the measuring
time $t_{ms}=\delta q^2/(\kappa e)^2$.

During the measurement, the voltage fluctuations on the SET island
will induce transitions between the qubit states. The rate for
excitation/relaxation of the qubit is proportional to the
SET noise spectral density $S_V(\omega)$ at the negative/positive frequency
corresponding to the transition $\Delta E/\hbar$.
The information of the initial state is destroyed at the
combined rate~\cite{MakhlinRMP,DevoretNature}
\begin{equation}
\label{mixing_rate}
\Gamma_1=\frac{1}{t_{mix}}=\frac{e^2}{\hbar^2}\kappa^2
\frac{E_J^2}{\Delta E^2}
\left[S_V(\frac{\Delta E}{\hbar})+S_V(-\frac{\Delta E}{\hbar})\right],
\end{equation}
here given in the relevant limit $\Delta E \gg E_J$ used below.

The spectral density of voltage fluctuations on the SET island is
described by the Fourier transform of the voltage-voltage correlation
function
\begin{equation}
\label{noisedef}
S_V(\omega) = \frac{e^2}{C^2} \int_{-\infty}^\infty d\tau
e^{-i\omega\tau} Tr\{\rho_{st}(t_0) \hat{N}(\tau) \hat{N}(0)\} .
\end{equation}
Here $\rho_{st}(t_0)$ is the density matrix of the system
in steady-state, which is assumed to have been reached
at some time $t_0$ before the fluctuation occurs ($t_0 < \min\{0,\tau\}$).
$\rho_{st}$ is the tensor product of the equilibrium (Fermi distributed)
density matrix $\rho^e_{eq}$ for the electron degrees of freedom in each
reservoir ($L, R, I$) and a reduced density matrix $\rho^c_{st}$,
describing the charge degrees of freedom, which we assume to be
diagonal~\cite{SchoellerNATO} with elements $P^{st}_N$ denoting the
probability of being in charge state $N$.

\begin{figure}
\includegraphics[width=8cm]{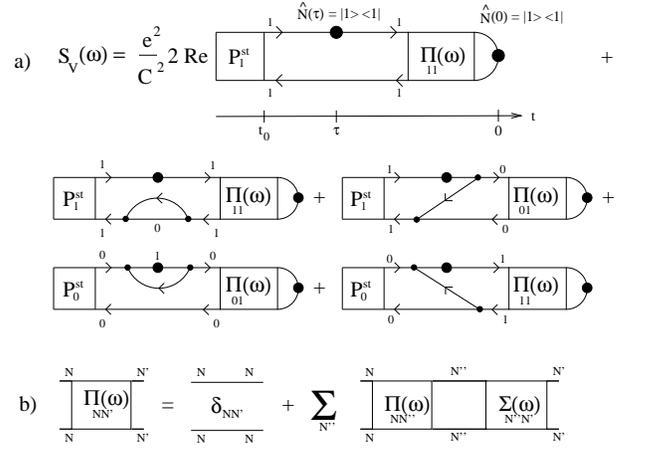}%
\caption{
a) Diagrammatic expression for $S_V(\omega)$ in our approximation,
neglecting processes involving other charge states
than $N\in\{0,1\}$. 
b) Dyson-type equation for $\Pi_{NN'}(\omega)$,
the frequency-dependent transition rate
between charge states $N$ and $N'$. }
\label{diag_fig}
\end{figure}

To evaluate $S_V(\omega)$ we make a perturbation expansion
of the forward and backward time evolution operators in terms of
the tunneling term $H_T$, as shown in Fig.~\ref{diag_fig}a.
The reservoir degrees of freedom are traced out using Wick's
theorem. In the diagrammatic language of reference \cite{SchoellerPRB}
(Fig.~\ref{diag_fig}) we have forward and
backward propagators of the Keldysh contour (horizontal lines),
with internal charge transfer vertices $e^{\pm\hat{\Phi}}$ (small dots)
connected by reservoir lines. The $\hat{N}(0)$ and $\hat{N}(\tau)$
operators form external vertices (big dots).
Figure~\ref{diag_fig}a shows the diagrammatic expression for
$S_V(\omega)$ in our approximation, neglecting processes
involving other charge states than $N\in\{0,1\}$.
Here $\Pi_{NN'}(\omega)$ is the frequency dependent rate
for transitions between the charge states $N$ and $N'$.
One may graphically write down a Dyson type of equation for
$\Pi_{NN'}(\omega)$ (Fig.~\ref{diag_fig}b).
In matrix notation this reads
\begin{equation}
\label{pi_omega}
\hat{\Pi}(\omega)=\hat{1}\frac{i}{\omega}+
\hat{\Pi}(\omega)\frac{i}{\omega}\frac{i\hat{\Sigma}(\omega)}{\omega}
\ \Rightarrow \ 
\hat{\Pi}(\omega)=\frac{i}{\omega}
\left[\hat{1}-\frac{i}{\omega}\hat{\Sigma}(\omega)\right]^{-1},
\end{equation}
where $\Sigma_{NN'}(\omega)$ represents the sum of irreducible
diagrams, i.e. containing no free propagators, for transitions
between $N$ and $N'$.

\begin{figure}
\includegraphics[width=8cm]{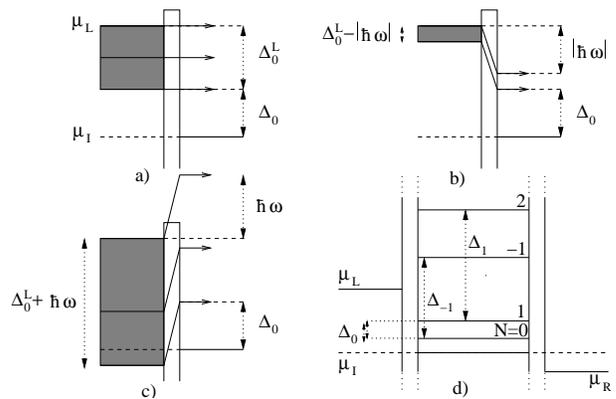}%
\caption{
a)-c) Schematic illustration of the frequency dependence of
the rate for tunneling from the left lead to the island,
when the island is in the $N=0$ state. The shaded area
indicates the amount of electrons energetically allowed
to tunnel; 
a) $\hbar\omega=0$,
b) $\hbar\omega < 0$ (SET emits energy) and
c) $\hbar\omega > 0$ (SET absorbs energy).
d) Schematic figure showing the bias used in the discussion
of the noise at different frequencies.
}
\label{proc_fig}
\end{figure}

Since the SET is biased in transport mode, 
we consider only the dominating lowest order
diagrams in $\Sigma_{NN'}(\omega)$, corresponding to
single tunneling events.
In this approximation we get \cite{SchoellerPRB}
\begin{equation}
\Sigma_{N,N\pm1}(\omega)=\gamma^\pm_N(\omega)+\gamma^\pm_N(-\omega),
\end{equation}
where $\gamma^\pm_N(\omega)$ denotes
the rate, in second order pertubation theory,
to go from charge state $N$ to $N\pm 1$ while the SET is
absorbing ($\omega$) or emitting ($-\omega$) a quantum 
of energy $|\hbar\omega|$.
The well-known expressions for these rates are  
\begin{equation}
\gamma^+_N(\omega)=\frac{\pi}{\hbar} \sum_{r}
(\hbar\omega+\Delta_N^r)\ \alpha_0^r\ n(\hbar\omega+\Delta_N^r),
\end{equation}
\begin{equation}
\gamma^-_N(\omega)=\frac{\pi}{\hbar} \sum_{r}
(\hbar\omega-\Delta_{N-1}^r)\ \alpha_0^r\ n(\hbar\omega-\Delta_{N-1}^r)
\end{equation}
where the dimensionless conductivity
\begin{equation}
\alpha_0^r=\sum_n \left|T^{rn}\right|^2\rho_r^n \rho_I^n
=\frac{R_K}{4\pi^2 R^r_T}
\end{equation}
is the ratio of the quantum resistance $R_K=h/e^2$
and the resistance of barrier $r\in\{L,R\}$, and
$\beta=1/k_B T$. $n(E)=1/\{1-\exp[-\beta E]\}$
is the Bose-Einstein distribution of electron-hole excitations and
comes from the convolution of the Fermi distributions for
filled initial states and empty final states.
$\Delta^r_N=V(N)+\mu_r-V(N+1)-\mu_I$
is the energy gained by an electron tunneling from the chemical
potential of lead $r$ ($\mu_r$) to the chemical potential of
the island ($\mu_I$), thus changing the charge state from 
$N$ to $N+1$ (see Figs.~\ref{proc_fig}a-c).
We also assume that the relaxation processes on the island
are fast on the timescales we are looking at and
neglect the energy dependence of both
tunneling matrix elements $T^{rn}_{kl} \approx T^{rn}$ and
reservoir densities of states
$\rho_r^n(\epsilon), \rho_I^n(\epsilon)$.
To avoid renormalization effects~\cite{SchoellerPRB} we
choose not too small bias,
i.e. $\max_r \alpha_0^r ln(E_c/|\Delta^r_0|)\ll 1$.

%
We also choose the bias not too high, so that the
higher charge states are inaccessible at low frequency,
i.e. $\gamma_1^+(0)=\gamma_0^-(0)=0$ giving $P_N^{st}=0$ for
$N\notin\{0,1\}$.
At low frequencies we may thus restrict ourselves to the two
lowest charge states $N\in\{0,1\}$. The matrix inverse
in Eq.~(\ref{pi_omega}) may then easily be performed analytically.
For high enough frequencies also the charge states
$N\in\{-1,2\}$ are accessible. In this regime
we may expand Eq.~(\ref{pi_omega}) to first order
in $\hat{\Sigma}/\omega$. The results in both these regimes
combine to the following expression for the noise
valid, within our approximations, for all frequencies:
\begin{equation}
\label{all_freq_noise}
S_V(\omega)=\frac{2e^2}{C^2}
\frac{P^{st}_0 \left[\gamma_0^+(\omega)+
\gamma_0^-(\omega)\right] +
P^{st}_1 \left[\gamma_1^-(\omega) +
\gamma_1^+(\omega)\right]}
{\omega^2+\left[\gamma_0^+(\omega)+\gamma_1^-(\omega)
+\gamma_0^+(-\omega)+\gamma_1^-(-\omega)\right]^2},
\end{equation}
where
$P^{st}_0=\gamma_1^-(0)/(\gamma_0^+(0)+\gamma_1^-(0))$ and 
$P^{st}_1=1-P^{st}_0$.

Note the simple structure of this expression: a sum over
the probabilities of being in the states $N=0$ and $N=1$ times the rates
for possible transitions from these
states, normalized by a denominator containing their finite
lifetime due to the bias.

At zero frequency Eq.~(\ref{all_freq_noise}) coincides with
the classical shot noise result
\cite{Korotkov_first,Korotkov_spectral,DevoretNature},
which is recovered by approximating all rates with their
zero frequency value.
In the high-frequency limit,
where the biasing energy is negligible, one recovers the
high-frequency Nyquist noise from two resistances connected
in parallell to ground.

It is informative to analyze Eq.~(\ref{all_freq_noise}) at zero temperature
where the tunneling rates are linear functions of the possible
energy gain (since $n(E)=\Theta(E)$, where $\Theta(x)$ is the
unit step function).
To be specific we assume that $\mu_L > \mu_I > \mu_R$
and $\Delta_0 > 0$ so that electrons tunnel from the left
lead to the right and $N=0$ is the lowest charge state.
We also assume that the left bias
is larger than the right one, $|\Delta_0^L|>|\Delta_0^R|$
(see Fig.~\ref{proc_fig}a).

\begin{figure}
\includegraphics[width=8cm]{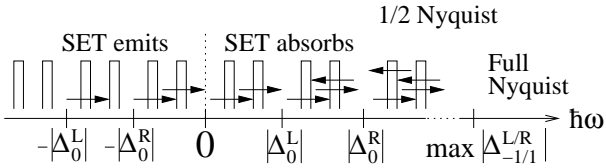}%
\caption{Schematic picture of the processes contributing
to the noise in different frequency regimes. Each arrow
(passage through a barrier) involves emission or
absorption of $|\hbar\omega|$.
}
\label{freq_proc}
\end{figure}

For large negative frequencies $\hbar\omega < -|\Delta_0^L|$
the fluctuations try to extract more energy than the bias can provide,
thus there is no contribution to the noise from this
frequency range (since we neglect higher order cotunneling effects).
Between  $-|\Delta_0^L| < \hbar\omega < -|\Delta_0^R|$ the SET may
emit energy $|\hbar\omega|$ while tunneling through the left
barrier, while for $-|\Delta_0^R| < \hbar\omega < 0$ tunneling through
both barriers contribute.
For positive frequencies the SET may absorb energy $\hbar\omega$ 
while tunneling through either barrier. For frequencies
$\hbar\omega > |\Delta_0^{L/R}|$ there is also a contribution
from electrons tunneling backwards, against the bias,
at the (L/R) junction. 
In the region $|\Delta_0^{L/R}| \ll \hbar\omega < 
\{|\Delta_{-1}^{L/R}|,|\Delta_1^{L/R}|\}$
the higher charge states $N\in\{-1,2\}$ are still inaccessible and the noise
approaches exactly one half of the Nyquist noise. This may be explained
by the strong correlation of fluctuations at the two barriers induced
by the Coulomb interaction.
At high frequencies
$\hbar\omega > \{|\Delta_{-1}^{L/R}|,|\Delta_1^{L/R}|\}$ 
also processes at barrier $L/R$ involving the charge states
$N\in\{-1,2\}$ contribute, giving the real high-frequency limit of
Nyquist noise from two completely uncorrelated tunnel junctions.
Figure~\ref{freq_proc} illustrates the above discussion of which
processes contribute in which frequency region.

\begin{figure}
\includegraphics[width=8cm]{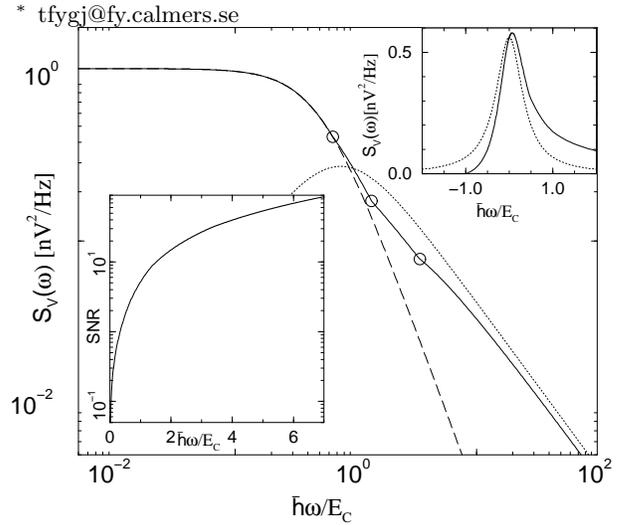}%
\caption{Summed voltage noise $S_V(\omega)+S_V(-\omega)$ for
a symmetrically biased SET with DC-current $6.7 nA$ and
$R_t^L=R_t^R=22 k\Omega$, $n_x=0.49$, $T=20 mK$,
$E_C/k_B=2.5 K$, Full expression (solid),
Classical shot noise (dashed), Nyquist noise (dotted).
The circles denote, in order from left to right, 
$\hbar\omega=\{|\Delta^L_1|=|\Delta^R_{-1}|,|\Delta^R_0|=|\Delta^L_0|,
|\Delta^R_1|=|\Delta^L_{-1}|\}$.
The right inset shows the low frequency regime separating positive
and negative frequencies (solid) compared with the classical
symmetrical expression (dotted).
The left inset shows the SNR of a single shot
read-out as a function of level splitting in the qubit.
}
\label{SV_fig}
\end{figure}

In the low frequency regime $|\hbar\omega|<|\Delta_0^{R}|$,
where no backward processes are allowed, the noise
spectral density may be written as
\begin{equation}
S_V(\omega)=\frac{e^2}{C^2}\frac{2I/e + 
2\pi\omega\left[P_0^{st}\alpha_0^L+P_1^{st}\alpha_0^R\right]}
{\omega^2+4\left[\gamma_0^+(0)+\gamma_1^-(0)\right]^2},
\end{equation}
where $I=2e\gamma_0^+(0)\gamma_1^-(0)/(\gamma_0^+(0)+\gamma_1^-(0))$
is the DC current through the SET. This is the expression for
classical shot noise plus a linear $\omega$-term in the numerator,
indicating that the probability of processes where the SET absorbs/emits
energy ($\pm\omega$) is raised/lowered compared to the classical expression
(see Figs.~\ref{proc_fig}c,d and the right inset of Fig.~\ref{SV_fig}).
The total frequency dependent decay rate of the states, found in
the denominator, is constant
in this regime since the linear increase/decrease in rates
for processes absorbing/emitting energy $|\hbar\omega|$ cancel
exactly, at zero temperature.

%
The sum of positive and negative frequency SET noise,
proportional to the back-action on the qubit, is plotted
in Fig.~\ref{SV_fig}, using the parameters of the RF-SET in
Ref.~\cite{Aassime} with sensitivity $\delta q=6.3\mu e/\sqrt{Hz}$.
One may define the signal-to-noise ratio ($SNR$) of a single shot
measurement as $SNR=\sqrt{t_{mix}/t_{ms}}$, which is plotted
as a function of qubit level splitting $\Delta E$ in the left inset
of Fig.~\ref{SV_fig}.
Using realistic qubit parameters for an SCB made with the same
aluminum technique as the RF-SET, $E_{qb}/k_B=1 K$, $\Delta E=2.4 E_{qb}$,
$E_J/E_{qb}=0.1$, assuming read-out at $\Delta E/E_{qb}=2.4$
we find that $SNR=\sqrt{t_{mix}/t_{ms}}\approx 5$,
which indicates that single-shot measurement is possible.
If one increases the qubit charging energy, e.g. by using a
niobium qubit, the $SNR$ increases even further.

In conclusion, we have obtained a simple expression for the
finite frequency voltage fluctuations of a SET island biased
in the transport mode. Using this information
about back-action we can conclude that single shot read-out
of a SCB qubit using an RF-SET is possible.
%



%
%

%

\begin{acknowledgments}
We would like to acknowledge fruitful discussions with
Per Delsing, Yuriy Makhlin, Vitaly Shumeiko, Tomas L\"ofwander
and Abdel Aassime.
We are especially grateful to Michel Devoret and Daniel Esteve for
the formulation of the problem.
This work was supported by NFR and the European Union
under the SQUBIT project.
\end{acknowledgments}


\end{document}